\documentclass[conference]{IEEEtran}
\usepackage{balance}

\usepackage{todonotes}

\usepackage{textcomp}

\usepackage{amsmath}

\ifCLASSINFOpdf
\else
\fi
\usepackage{url}

\begin{document}
%
\title{Analyzing Attacks on Cooperative Adaptive Cruise Control (CACC)}

\author{\IEEEauthorblockN{Rens van der Heijden, Thomas Lukaseder, Frank Kargl}
\IEEEauthorblockA{Institute of Distributed Systems\\
Ulm University, Germany\\
\url{rens.vanderheijden@uni-ulm.de}, \url{thomas.lukaseder@uni-ulm.de}, \url{frank.kargl@uni-ulm.de}
}
}

%


\IEEEoverridecommandlockouts
\IEEEpubid{\makebox[\columnwidth]{\copyright2017
IEEE \hfill accepted for VNC 2017 (no DOI available yet)} \hspace{\columnsep}\makebox[\columnwidth]{ }}

%

\maketitle

\begin{abstract}
  Cooperative Adaptive Cruise Control (CACC) is one of the driving applications of vehicular ad-hoc networks (VANETs) and promises to bring more efficient and faster transportation through cooperative behavior between vehicles.
  In CACC, vehicles exchange information, which is relied on to partially automate driving; however, this reliance on cooperation requires resilience against attacks and other forms of misbehavior.
  In this paper, we propose a rigorous attacker model and an evaluation framework for this resilience by quantifying the attack impact, providing the necessary tools to compare controller resilience and attack effectiveness simultaneously.
  Although there are significant differences between the resilience of the three analyzed controllers, we show that each can be attacked effectively and easily through either jamming or data injection.
  Our results suggest a combination of misbehavior detection and resilient control algorithms with graceful degradation are necessary ingredients for secure and safe platoons.
\end{abstract}


%

\section{Introduction}
\label{sec:intro}
\begin{figure}
\includegraphics[width=\columnwidth]{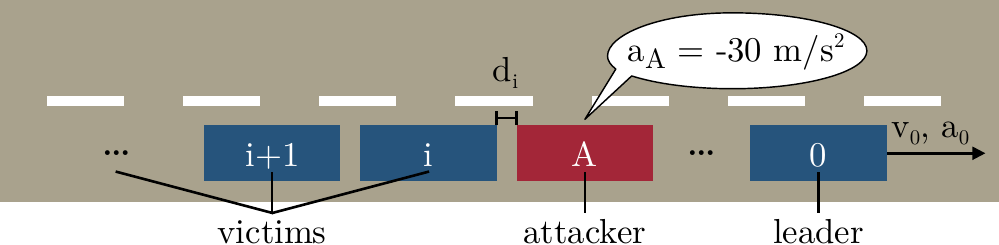}
\caption{This figure shows a platoon with a leader and several followers: in our work, the red vehicle is the attacker, and the platoon length is 8. The attacker transmits a malicious beacon with a false acceleration of $-30$m/s$^2$.}
\label{fig:platoon-example}
\end{figure}

In this paper, we study cooperative adaptive cruise control (CACC), an application of Vehicular Ad-Hoc Networks (VANETs) aimed at increasing road efficiency by partially automating driving.
CACC is essentially an extension of existing cruise control (CC) and adaptive cruise control (ACC) technologies, both of which are designed to allow the driver to maintain constant speed while driving.
In the case of ACC, vehicular sensors, such as RADAR, LIDAR and cameras, are used to measure the distance to the preceding vehicle, in order to automatically respond to changes in its' driving behavior.
CACC extends this concept by allowing vehicles to communicate and create a \emph{platoon}, consisting of a leader vehicle and multiple followers, as shown in Figure \ref{fig:platoon-example}.
There are many different controller implementations in the literature that enable this behavior~\cite{CSUR-Platooning} since the proposal of CACC proposal in the 90s.
It has been shown that CACC can be more efficient than ACC, and that even a constant spacing between vehicles (i.e., independent of platoon speed) can be theoretically achieved, whereas it has been shown that ACC alone cannot achieve this goal~\cite{Rajamani-CC-book}.

Since CACC was proposed, VANETs have made significant leaps in development, including the DSRC standard and a variety of ETSI, IEEE and SAE standards that define the various network layers.
In parallel to these developments, security standards were designed, in order to protect against attacks common in other fields.
In relatively early research, it became clear that integrity is the most significant security goal for VANETs in general~\cite{RayaSecuritySurvey}, although availability is also significant.
The IEEE 1609.2 standard defines how to protect against common external attacks, by providing message integrity and message authenticity (i.e., only legitimate vehicles can transmit messages).
Both the C2C-CC in Europe and the CAMP in the US have since started deployment of security credential management systems (SCMSs) to manage what constitutes legitimate vehicles.
In parallel, one of the major advances in the security community was the introduction of misbehavior detection, as a mean to detect and revoke vehicles that transmit malicious data.
This type of behavior is at the focus of this work, particularly because the desired increased road usage with CACC requires vehicles to move in a way that would be unstable if only ACC was deployed.

The research in misbehavior detection for VANETs concentrates mostly on being able to detect incorrect information in beacon messages.
These messages, defined by the previously mentioned standards, are periodically sent to all neighboring vehicles to inform them of the state of the sender.
One of the main application areas studied to evaluate such detection schemes is position-based routing, where multi-hop communication is enabled by finding routes based on vehicle positions.
Because vehicle position is also used in order to avoid potential accidents in normal traffic (e.g., by warning the driver with an alarm tone), many detection algorithms operate on position information specifically.
However, CACC typically uses acceleration and speed information to compute the behavior of the local vehicle based on that of the leader and predecessor vehicles (and in some cases, the entire platoon).
If this information becomes unavailable or is incorrect, accidents may occur.
The data for CACC controllers is essential to protect, but the security of these controllers when considering malicious insiders has only been studied by few authors~\cite{Dadras-Adversarial,DeBruhl-Misbehavior,Amoozadeh-Commag,Alipour-Jamming}.

Existing work shows, that attacks can impact safety (i.e., potentially cause accidents), but lack a rigorous analysis and a framework within which such attacks and the resilience of the respective controllers can be analyzed.
In this paper, we lay the groundwork for such a framework: we implement several attacks and provide a concrete quantification to measure the impact of an attack.
Going beyond simply finding whether an accident occurs or not, we analyze the significance of such an attack (in terms of the $\Delta v$ of the accident) and quantify how the stability of an individual platoon is affected even if no accident occurs.
We present results for three existing controllers implemented in Plexe~\cite{Segata-Plexe}.
Because we use this extensible platooning simulator as a basis for our evaluation, our work can easily be extended with new controllers.
Similarly, our attack implementation can easily be extended to study the impact of more attacks.
We publish the source code, data analysis scripts, and recorded data in order to stimulate further research in this area.

The remainder of this paper is organized as follows.
Section \ref{sec:related-work} discusses the existing analyses of attacks on CACC and platooning, as well as some basics regarding the controllers in Plexe.
In Section \ref{sec:attacker-model} we divide existing attacks into three categories, of which two are analyzed extensively in this paper.
We also discuss which proposed attacks are infeasible under the assumption that current security standards are implemented correctly.
Section \ref{sec:analysis} describes our simulation setup and the corresponding evaluation methodology.
We discuss these results and their implications for CACC in Section \ref{sec:discussion}, and close with a conclusion in Section \ref{sec:conclusion}.

\section{Related Work}
\label{sec:related-work}
We distinguish two areas of related work: controllers, which implement the functionality that CACC and ACC offer, and existing studies on misbehavior in platooning and similar settings.

\subsection{Controllers}
\label{sec:controllers}
Platoons are vehicles driving in a group, attempting to keep a minimal but still safe distance between the vehicles within the group.
This can be achieved using ACC and CACC; the main reason for using CACC is that the safe distance can be much smaller.
In both settings, a platoon normally consists of a single leader and a set of followers, together referred to as the platoon (or sometimes, string).
The set of followers can be dynamic, although this usually requires some kind of join and leave procedures.
Each vehicle implements a controller that determines the desired acceleration, which is used to configure the vehicles' engine and brake controller.
In this work we focus on the upper controller (i.e., setting the desired acceleration) exclusively, since this is the controller that an attacker can easily reach (either through communication or through attacks on sensors).

There is a broad scope of literature from the control theory and engineering side covering a variety of different controllers~\cite{CSUR-Platooning}.
This literature typically introduces two stability metrics: local stability and string stability.
Here, local stability refers to a vehicles' distance to the next vehicle, while string stability is a property of the platoon that determines how spacing errors propagate through the platoon.
A platoon is considered string stable when the $\infty$-norm of the transfer function is less than $1$ ($||\hat{H}(s)||_{\infty} \leq 1$)~\cite{Rajamani-CC-book}, which basically means that the platoon will eventually converge to a stable state if the leader keeps a specific velocity perfectly.
Maintaining the distance to the next vehicle can done by keeping a fixed distance (i.e., meters) or by maintaining a speed-dependent headway time (i.e., seconds), referred to as constant spacing and constant-time spacing.
Earlier work has shown that non-cooperative methods (i.e., ACC) cannot achieve string stability in a constant spacing policy~\cite{Rajamani-CC-book}.

In this work we focus on controllers implemented by Segata~et~al.~\cite{Segata-Plexe} in Plexe, which includes a constant spacing controller, a controller with graceful degradation and a consensus controller.
The constant spacing controller is taken from Rajamani~\cite{Rajamani-CC-book} and keeps a fixed distance based on the measured distance and the received speed and acceleration of the preceding vehicle and the leader vehicle.
The controller by Ploeg~et~al.~\cite{Ploeg-Platooning} is designed to enable graceful degradation on network errors: their degraded CACC controller predicts the acceleration of the preceding vehicle if this information is not received.
The idea is that this degraded controller outperforms ACC, and can thus be used to bridge communication gaps; this is particularly interesting for jamming attacks.
The last controller is a consensus algorithm by di~Bernardo~et~al.~\cite{diBernardo-Consensus}, which uses the position, speed, and acceleration information of all vehicles in the platoon.
This controller has a larger spacing, but aims to maintain stability bidirectionally (i.e., vehicles also adapt their behavior to that of vehicles behind them); this is particularly interesting with respect to data injection attacks.

\subsection{Existing Attacks}
\label{sec:attackers}

Attacks on CACC have typically been studied from two different perspectives, mostly working in parallel.
Security research in vehicular networks has already suggested new types of attacks for the overall network in the previous decade~\cite{RayaSecuritySurvey}.
More recent work has concentrated on misbehavior detection~\cite{vanderHeijden2016} and data injection attacks, which are uniquely suitable for vehicular networks and other cyber-physical systems.
Platooning is one of the motivating examples for these researchers.
On the other hand, authors from the control theory side have gained an understanding of potential security concerns, and have started developing attack-resilient control algorithms.
In terms of attacks, some authors have also investigated how to exploit knowledge of the controller to show that attacks causing instability are possible.

Alipour-Fanid~et~al.~\cite{Alipour-Jamming} discussed platoon behavior under jamming. In their attack scenario, the jammer is a drone flying above the platoon, aiming to use its' limited power to disrupt the platoon from above.
Jamming is thus done reactively (i.e., with success rate $<100\%$) to conserve power, although power consumption is not considered further.
Their work shows that their ACC controller is string stable at 2.2 seconds headway distance, while CACC under the influence of their reactive jammer is string stable at 1.7 seconds headway distance.
Additionally, they show the attack works best when the attacker is located near the first vehicle in the platoon.

Amoozadeh~et~al.~\cite{Amoozadeh-Commag} discuss a two-level controller model and describe a variety of attacks, including application and network layer attacks, as well as other issues such as sensor tampering and privacy issues.
They present a result of a message falsification attack, which in their setting is a vehicle external to the platoon that falsifies all beacons; the result of the attack is platoon instability (i.e., the platoon does not converge back to a stable state).
They also explain that such an attack is most effective when acceleration changes occur.
Similar results are presented for a jamming attack -- here, the authors' controller downgrades to ACC automatically.
In our work, we aim to show that such a downgrade is not necessarily sufficient, depending on how system parameters are chosen.

DeBruhl~et~al.~\cite{DeBruhl-Misbehavior} also use a two component controller, consisting of a PD feedback controller (radar-based) to keep distance and a feedforward controller (communication-based), whose outputs are added together to form the desired acceleration that is provided to the vehicle drivetrain.
This controller has been shown to be string-stable in real networked platoons, and DeBruhl~et~al. examine its' behavior under various attack strategies, including a collision induction attack.
The authors continue to develop an error calculation and detection algorithm, which essentially estimates the expected behavior of the vehicle in front and switches to ACC if this vehicle appears to behave differently than it claims to.

Outside of the context of communication, but also relevant to secure and safe platooning, is the work from the research group of Gerdes~et~al.~\cite{Gerdes-Efficiency} and Dadras~et~al.~\cite{Dadras-Adversarial}.
These authors looked at ACC-based platoons, and examined adversarial behavior within such a platoon, where the attacker systematically disrupts the distributed control algorithm using another algorithm that manipulates control input of their own vehicle to achieve their goal.
In the first work, the authors show that for a given PID controller, they can cause significantly increased energy consumption in neighboring vehicles; as reduced energy consumption is one of the main drivers of CACC research, this impact is significant.
The second work shows that for a PD controller, the attacker can cause the platoon to be asymptotically unstable, i.e., destabilize and eventually dissolve the platoon.

\section{Attacker Model}
\label{sec:attacker-model}
In this section, we describe our attacker model, which covers attacks proposed in the literature above, as well as elements from the area of misbehavior detection~\cite{vanderHeijden2016} and attacks on sensors~\cite{Petit2015}.
In particular, we are interested in how realistic the attacks proposed in the literature are, when considering currently standardized security mechanisms.
Standardized security mechanisms include the use of pseudonymous certificates to protect driver privacy, the corresponding private keys of which are stored in a hardware security module (HSM), which is somewhat protected against tampering.
The core of our attacker model is that an attacker may gain control over a vehicle, and thus over legitimate key material.
This can happen either through software compromise (e.g., malware spread through applications running on the entertainment system), by physical manipulation of the vehicle, or through key extraction from an old HSM.
However, unlike traditional network attacker models, the attacker additionally must obey the laws of physics that are observable by any network participant (e.g., an attacker is not omnipresent throughout the network).

For CACC specifically, we assume the software and protocol implementations are suitable, i.e., the attacker cannot find an exploit here, and must obey the protocol.
In other words, we restrict the target of the attack to the platooning application, and assume other mechanisms are in place to filter out invalid or improperly formatted messages.
In our model, the attackers' primary goal is to cause a crash within the platoon, with the secondary goal of destabilizing it.
We assume the attacker is already part of the platoon, a position from which the attacks can do the most damage.
However, as the literature points out~\cite{DeBruhl-Misbehavior}, if the attacker controls the leader vehicle, they have complete control over the platoon, and therefore attacks should be performed by a non-leader vehicle.
We follow this argument and also assume attacks are executed by a follower in order to study the potential impact of attacks.
Another argument for this is that leader vehicles can be placed under additional scrutiny and replaced if necessary, while followers are more difficult to replace if they are non-cooperative.

\subsection{Jamming}
Denial of service (DoS) attacks includes a variety of jamming attacks (regular and reactive), which are the most popular variant of this attack within vehicular networks.
This also includes various attacks on higher layers that influence reception (e.g., violating the MAC protocol, causing others to wait).
Some authors~\cite{Amoozadeh-Commag} also suggest a DoS attack may be launched on the network layer, in order to disrupt communication by overloading the hardware security module (HSM).
This leads to a form of \emph{cryptographic packet loss}, i.e., packet loss caused because messages cannot be verified, or cannot be verified in time~\cite{Feiri-CertOmission}.
In this work, we consider all of these attacks to have the same effect, i.e., that messages will no longer arrive, and we refer to it as jamming.
This generalizes the attack compared to the related work~\cite{Alipour-Jamming}, where the authors considered reactive jamming.

\subsection{V2X Data Injection}
The second class of attacks we identify is V2X Data Injection, which generalizes a variety of different attacks discussed in related work~\cite{Amoozadeh-Commag,DeBruhl-Misbehavior}.
Any attack that requires the attacker (or a vehicle compromised by the attacker) to send modified packets falls into this category; it is particularly useful because these types of attacks may be feasible to mitigate through misbehavior detection.
The specific implementation of this attack is usually dependent on the type of data exchanged by the controller, and thus the corresponding impact is different for every controller.
Similarly, the goal of an attacker is significant for the implementation, and sometimes used as a distinguishing factor in the literature (this includes the reduced headway attack, the collision induction attack and the misreporting attack, among others).
However, these goals can often be achieved through one of the other classes of attacks (e.g., collision induction is possible through jamming), and thus we avoid this terminology.

Spoofing and replay attacks can also be considered examples of this class, but these are considered mostly solved in the vehicular networking security community.
In vehicular networks, where GPS is often assumed to be practically ubiquitous, replay attacks can be protected against using (close to) synchronized clocks.
Similarly, because cryptographic keys are authenticated through pseudonymous certificates, and messages are protected by digital signatures, spoofing attacks that cause the receiver to misinterpret the origin of the message are not possible.
This also applies to message falsification attacks (i.e., attacks that manipulate messages of other vehicles), for which digital signatures are an adequate countermeasure.
However, it is possible to use multiple certificates to simulate more than one vehicle -- this is referred to as a Sybil attack (as originally introduced by Douceur~\cite{Douceur-Sybil}).
Whether this attack is feasible depends on how easy it is for the attacker to obtain pseudonymous certificates that can be used simultaneously.

\subsection{Sensor Manipulation}
The third and final class of attacks is sensor manipulation, which constitutes any attack that causes the internal network of the vehicle to report incorrect information to the controller.
An attacker could either have access to the internal network of the vehicle and transmit false data here, causing the vehicle to react incorrectly (or even destabilize the platoon through the controller attack~\cite{Dadras-Adversarial}).
Although this could also be done in software, i.e., manipulating memory before the controller processes the inputs, attacks on the internal network are considered to be more interesting, because misbehavior detection is a potential countermeasure.
Software attacks, on the other hand, are difficult to prevent through misbehavior detection, as the detection system will likely also be compromised in such a setting.

Another variant of sensor manipulation positions the attacker outside of the target vehicle: instead of attacking the network, the attacker uses their sensors and other tools to blind the sensors of the target vehicle, or cause them to behave in a certain way.
Researchers have shown that both LIDAR and cameras are vulnerable to such attacks~\cite{Petit2015}, and the effect of such an attack is very similar to the other type of sensor manipulation attack (i.e., the controller sees invalid inputs).
The only distinction is where the attacker is located, which influences the cost of such an attack, but the expected results for the platoon behavior are similar, and thus for the study of attacks we place these attacks in the same category.

\section{Attack Analysis}
\label{sec:analysis}
In this paper we focus on communication only: we want to answer the question how CACC behaves under misbehavior, i.e., what are the potential impacts of such attacks.
The aim of this work is to provide a general framework for attack analysis, which is helpful for the development of resilient control algorithms, as well as the misbehavior detection community.

\subsection{Simulation Setup}
\label{sec:setup}
We use Plexe (version 2.0)~\cite{Segata-Plexe}, which is a simulation toolkit based on VEINS, as basis for our analysis due to its' flexibility in implementing control algorithms and accurate simulation of network behavior.
Plexe also extends SUMO, which provides realistic microsimulations of physical vehicles; it extends this toolkit with various controllers: a constant spacing controller~\cite{Rajamani-CC-book}, a loss-tolerant controller~\cite{Ploeg-Platooning} and a consensus controller~\cite{diBernardo-Consensus}.
For our implementation, we leave Plexe mostly unchanged, and only extend it to include our attacks.
We implement four attacks: one jamming attack and three data injection attacks.

Our jamming attack is designed to analyze the maximum possible impact of an attack, and is therefore implemented by locally dropping received messages after the attack starts.
This simulates any type of denial of service, including cryptographic packet loss and various jamming strategies, and provides an upper bound on the attack success, although it may overestimate it in some cases.
The alternative would be to implement a specific attack (e.g., as done by Alipour-Fanid~et~al.~\cite{Alipour-Jamming}), but this approach is not as generalizable as ours.

\begin{table}
  \centering
  \caption{Excerpt of simulation parameters: full parameters available in the data and source code repositories.}
  \begin{tabular}{c|c}
    transmit power & 100mW \\
    sensitivity & -94dBm\\\hline
    platoon length & 8\\
    attacker ID & 3\\\hline
    controllers & CACC, Ploeg, Consensus\\
    CACC spacing & 5, 7, 9, 11, 13, 20 m\\\hline
    target speed & 50, 80, 100, 120, 150 km/h\\\hline
    attack: speed value & -50, 0, 50, 100, 150 m/s\\
    attack: accel value & -30, -10, 0, 10, 30 m/s$^2$\\
    attack: position shift & 3, 5, 7, 9, 11 m/s\\
  \end{tabular}
  \label{tab:simparams}
\end{table}

Our data injection attacks are all implemented by exchanging some data with false data in the message an attacker would send according to the normal control algorithm.
We analyze three different types: injecting false positions, false speeds or false acceleration values.
There are multiple ways to implement these attacks; in our case, we set the attackers' value to a constant value for speed and acceleration, which greatly simplifies the interpretation of the results.
In addition, we again aim to provide an upper bound on the attack effectiveness.
For position falsification, we provide a position error that increases linearly over time, and is added to the vehicles' current position for each beacon.

The simulation scenario we analyze is a standard scenario from Plexe, which was used by the authors to analyze platoon behavior: the sinusoidal scenario.
This scenario initializes the platoon and then provides the leader with a specific acceleration profile that leads to a sinusoidal speed graph.
By performing attacks at different points in time, we can use this scenario to approximate the impact of attacks on real behavior.
The significant simulation parameters can be found in \ref{tab:simparams}; the complete configuration file and code is available on Github\footnote{\url{https://github.com/vs-uulm/vnc2017-CACC-code}}.
Because of the way Plexe is implemented, simulating chain collisions reliably was not possible, so we collected information about the first collision only.

\subsection{Metrics}
\label{sec:metrics}
In this work, we aim to analyze the effectiveness of various attacks on different CACC controllers.
The goal of our attacker is to either cause a crash or maximize instability of the platoon.
As such, we distinguish between two cases: a crash or some instability.
If a crash occurs, impact velocity $\Delta v$ is used to quantify the impact of the attack (medical research suggests that this is the best predictive factor for injury~\cite{Joksch1993-Velocity}).
If there is no crash, we quantify the instability in the platoon for the remainder of the simulation (approximately 30 seconds).
Since all the studied controllers have been shown to be string stable, we can assume that the platoon eventually stabilizes.
The option we chose is $max_{s,i,t}(e)$, where $s$ is the simulation run, $i$ is the $i$th vehicle in the platoon, $t$ is time and $e$ is spacing error, which most accurately represent the worst case.
This was inspired by earlier work~\cite{Alipour-Jamming}, where a similar metric is used.
The spacing error is defined as the difference between the desired distance of the controller and the actual distance between two vehicles.
The alternative metrics were $avg_s(avg_i(max_t(e)))$, the average maximum spacing error, and $avg_s(avg_i(max_t(a)))$, the average maximum acceleration, both of which can be considered metrics for passenger comfort.
Usability studies have shown~\cite{Larburu2010-Safe} that spacing is important for user acceptance; we included the acceleration as an additional metric, which is also common in transportation research.
Averaging over simulations and vehicles allow us to estimate the average comfort of all vehicles in a platoon; however, the maximum spacing error more accurately represents the potential inefficiency and risk due to an attack, which is why we decided on this metric\footnote{The data repository for this paper includes all metrics.}.

\begin{figure}
\includegraphics[width=.85\columnwidth]{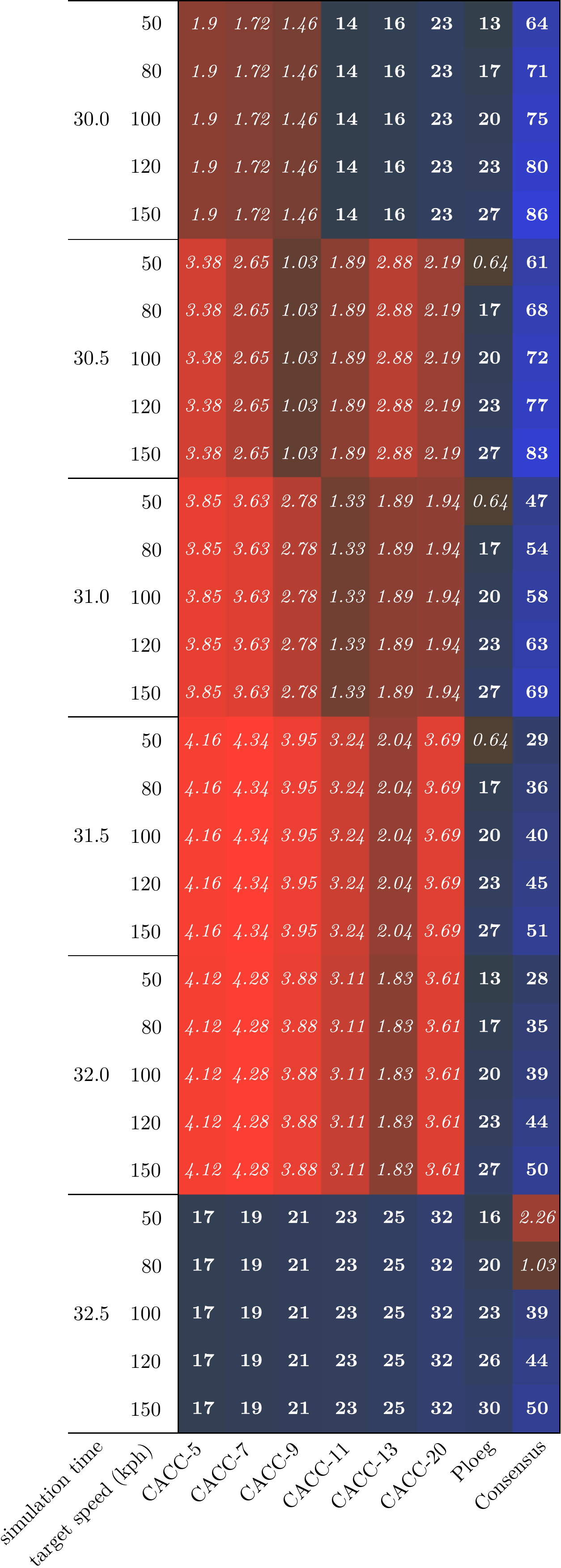}
\caption{Heat map showing attack success and impact for the constant jamming attack at the given point in simulation time, for different target platoon speeds and different controllers. Red/italics: attack leads to accident; number is impact velocity in m/s. Blue/bold: no accident; number is maximum spacing error in m.}\label{fig:jamming-heatmap}
\end{figure}

\subsection{Results}
\label{sec:results}
This section presents the results of our attack experiments, and shows how we used our metrics to gain insight into the effectiveness of different attacks, as well as the resilience of the existing CACC algorithms.
The data on which these results are based will also be published on Github\footnote{\url{https://github.com/vs-uulm/vnc2017-CACC-data}}, although they can also be re-generated using the source code and configuration mentioned above.

\subsubsection{Jamming}
We first discuss the results of our jamming attack, which are shown in Figure \ref{fig:jamming-heatmap}.
This figure includes both successful attacks (i.e., crashes, marked in red) and stability impact (i.e., no crash, marked in blue), as well as the corresponding impact (see Section \ref{sec:metrics}).
The figure shows aggregated values (using the previously discussed aggregation approach) over five runs. In all cases, the impact of the attack remained similar (i.e. all five runs resulted in collisions or all five runs did not result in collisions).
We simulated this for different \emph{target speeds} and different \emph{jamming times}: the target speed is the intended average speed of the platoon (i.e., the average speed of the leader vehicle), while the jamming time is the time at which jamming starts.
The jamming time directly corresponds to the range in which acceleration of the leader vehicle is positive; since the leader data is disseminated throughout the platoon, the attacker can use this acceleration profile to choose the time of attack.
Based on the speed profile generated from Plexe's standard configuration, we determined that the positive acceleration lies between 30 and 32.5 seconds of simulation time (since the behavior is sinusoidal, this pattern repeats every 5 seconds).

As can be seen in the figure, the target speed has no impact on the effectiveness of the attack on CACC. However, the attacks are more successful with higher velocities when Ploeg or Consensus are used as the spacing of these algorithms depends on the platoon target speed. We can therefore conclude, that spacing has a strong influence on attack effectiveness. On the other hand, attack effectiveness is not directly influenced by the target velocity. In general, with more spacing, attack impact decreases. The figure shows that this is not always true (e.g. CACC-13 vs. CACC-20 at 32s simulation time). This is due to the fact that with CACC-13 and lower, the attack occurs between the attacker (car 3) and the subsequent car 4. With CACC-20, car 4 can avoid colliding with the attacker, however, car 5 does collide with car 4.

\begin{figure}
\includegraphics[width=\columnwidth]{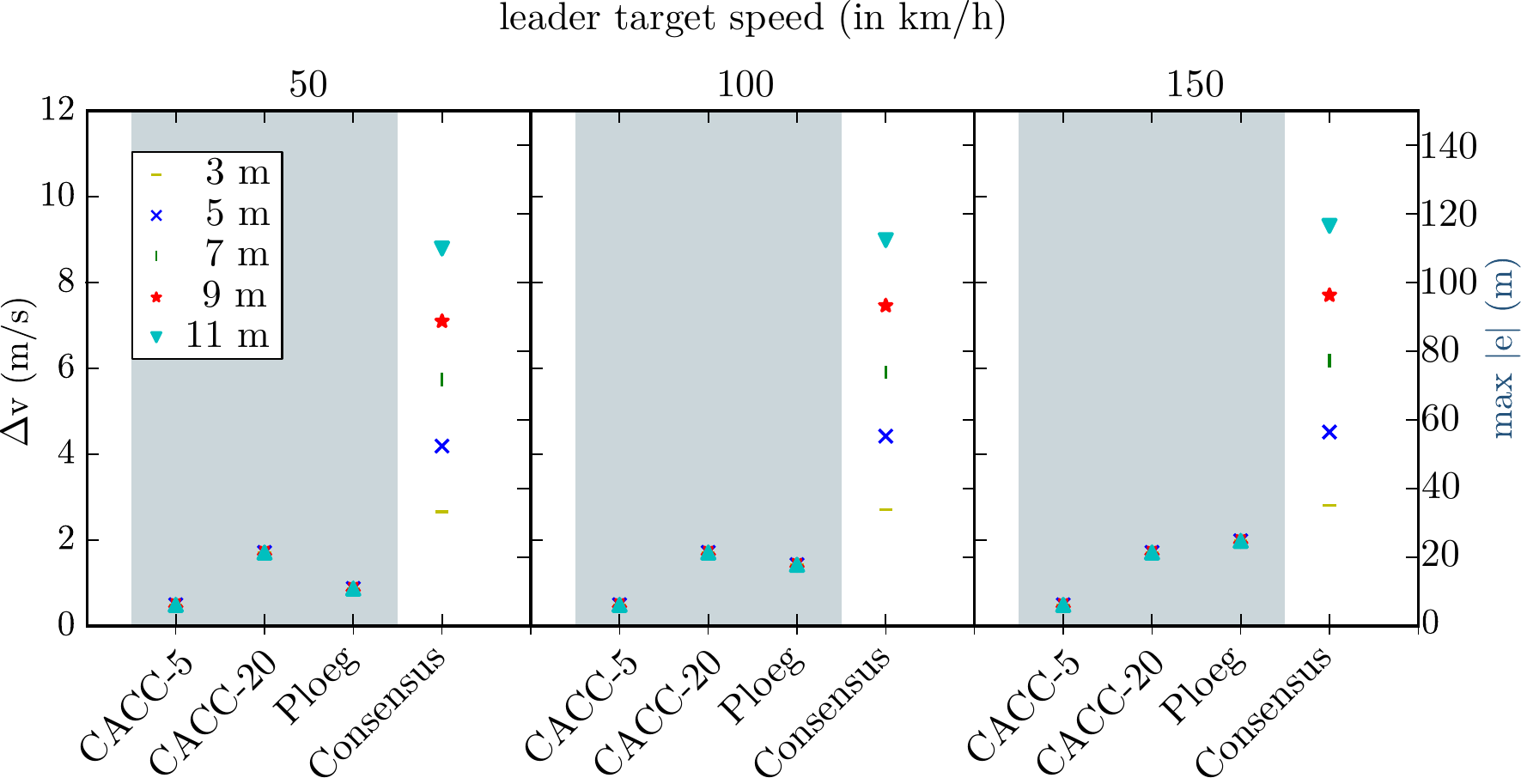}
\caption{The effect of position falsification: blue areas represent points not related to a crash.}\label{fig:posInjection}
\end{figure}

\subsubsection{Position Injection}
The next step in our analysis covers position falsification attacks, often cited as one of the most significant types of data injection attacks on vehicular networks.
In our model, the false position is a shift of the attackers' real position that increases over time.
It turns out that this attack is completely ineffective against the Ploeg and constant spacing controllers -- it only causes crashes when the consensus controller is used.
However, we can furthermore see that the spacing error of the Ploeg controller increases significantly with the leader target speed.
This is because the consensus controller is the only one that uses received position information directly.
Somewhat surprisingly, the impact of this attack on the consensus controller depends only on the value transmitted by the attacker, and not significantly on the target speed with which the leader (and thus the platoon) is moving.
The full results are shown in Figure \ref{fig:posInjection}: for each shift parameter (mark and color of the scatter plot), with different leader target speeds (top, left to right), the graph includes the results for four different controller configurations (bottom, left to right).
The impact speed of successful attacks (white background) varies from $2$ to $9$ m/s, while the maximum spacing error (gray background) is unaffected by the position falsification.
This means that all attacks that have any effect lead to an accident.

\begin{figure}
\includegraphics[width=\columnwidth]{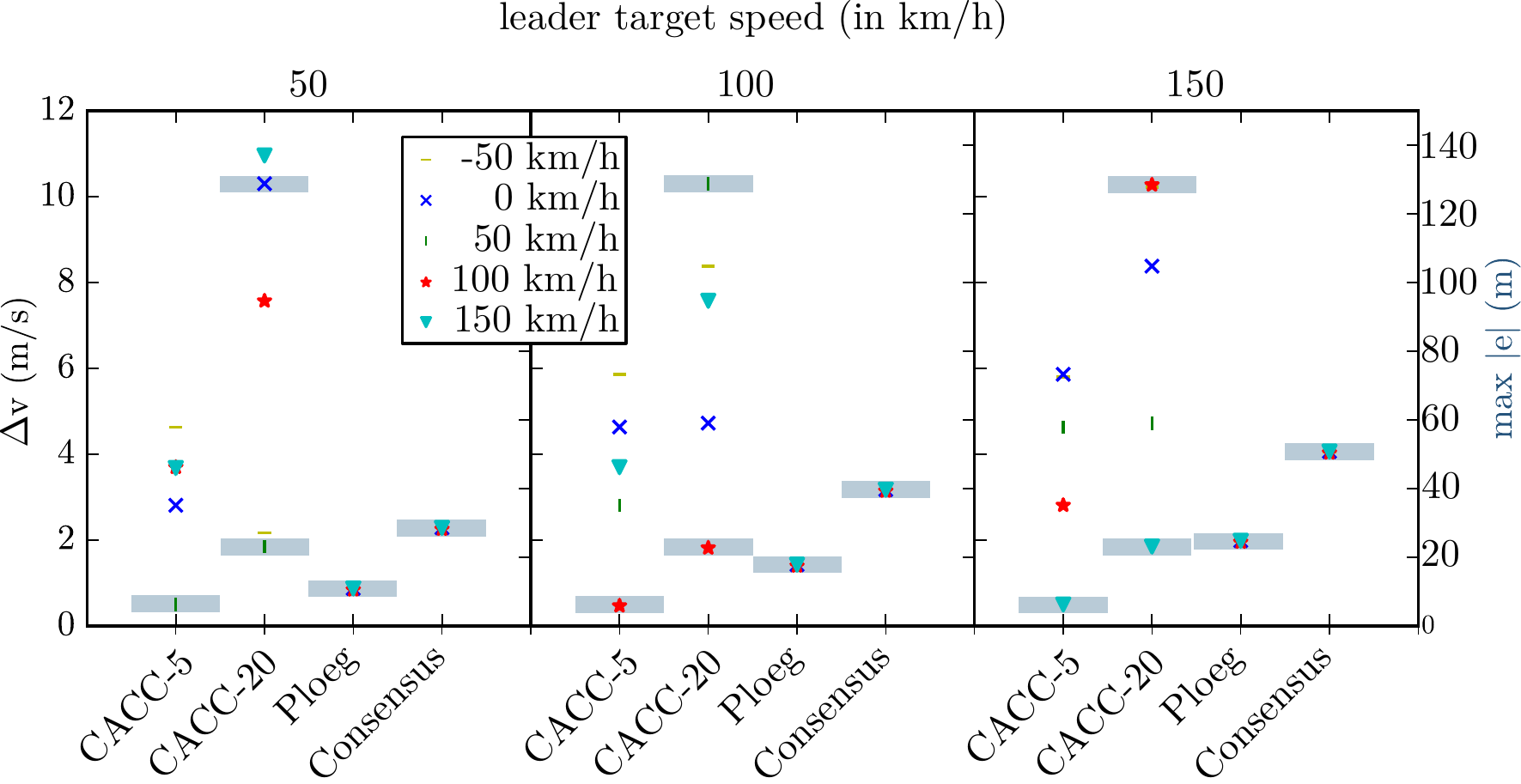}
\caption{Attacking with false speeds: blue areas represent points not related to a crash.}\label{fig:speedInjection}
\end{figure}

\subsubsection{Speed Injection}
Our next attack was manipulating the speed transmitted by the attacker.
Here, we continuously output the same value regardless of platoon or controller behavior, which causes the error between the attacker's value and actual value to vary from $10$ to $-10$ km/h if it falsifies the platoon target speed.
The result of our analysis, displayed in Figure~\ref{fig:speedInjection}, shows that speed injection is successful against constant spacing CACC only, where the most significant crashes happen at lower speeds; the attacker claims to accelerate, but actually does not.
When the attacker claims negative speeds (e.g., $-50$) the result is a high position error.
The Ploeg controller shows the same behavior as before (i.e., it is not affected by the attack), and here the consensus controller is also unaffected, although the spacing error is much higher.
Due to a constant falsified speed, the attacker does not achieve success in the constant value matching the leaders' speed; we expect this will change if the leader behavior is less predictable.

%

\subsubsection{Acceleration Injection}
\begin{figure}
  \vspace{0.006in}
\includegraphics[width=\columnwidth]{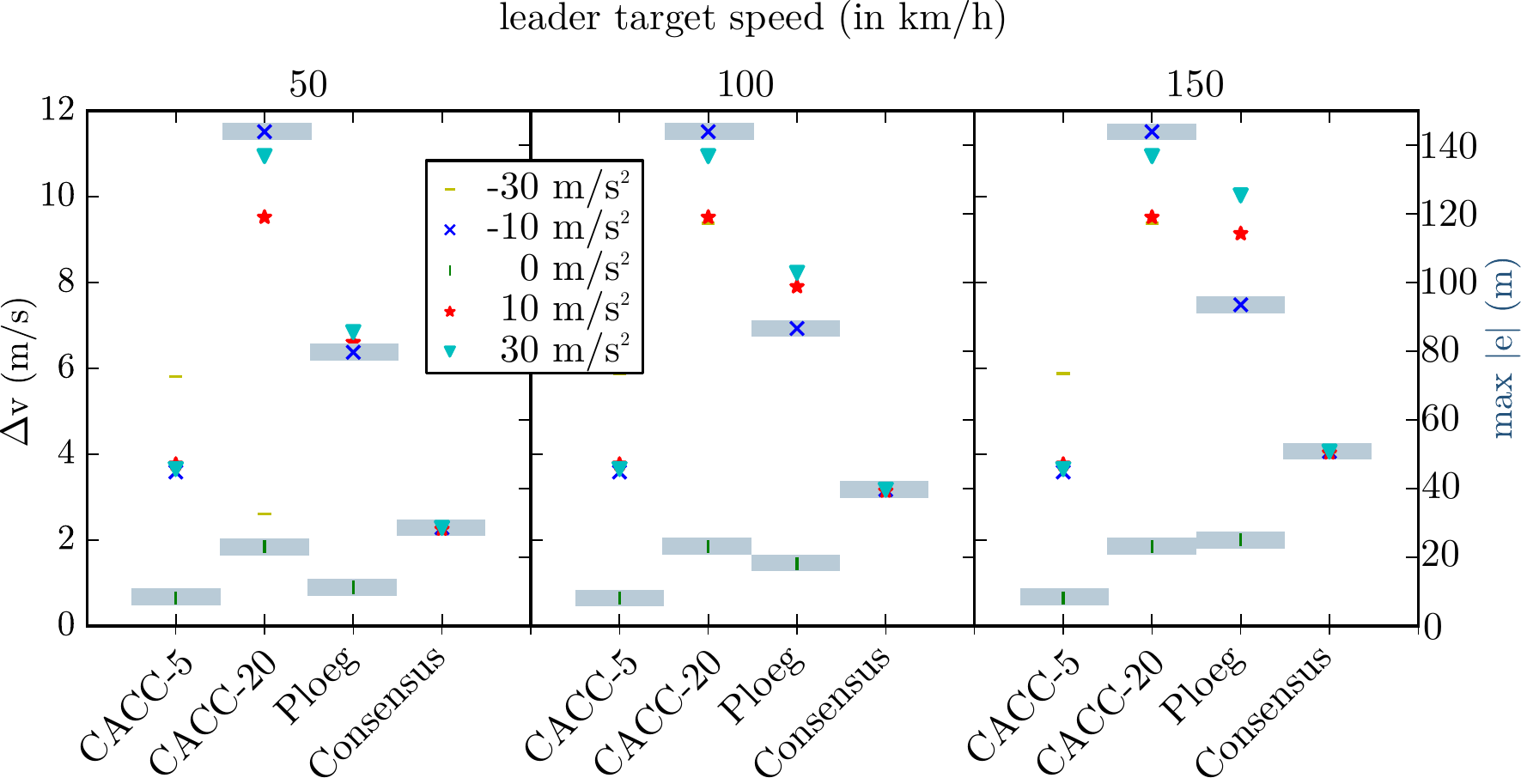}
\caption{Behavior of controllers with falsified acceleration information: blue areas represent points not related to a crash.}\label{fig:accInjection}
\end{figure}

The final attack we performed is manipulation of the transmitted acceleration.
Here, we also chose a relatively extreme scope, in order to discover the maximum potential of these attacks ($-30, -10, 0, 10, 30$ m/s$^2$).
The resulting statistics are shown in Figure~\ref{fig:accInjection}.
As expected, this attack affects all controllers, because acceleration is used in all the control algorithms.
However, the effect is widely different per controller: the consensus controller can tolerate these large errors in acceleration information (although the maximum spacing error increases slightly), while the others fail almost completely.
This is likely due to the fact that the consensus controller is the only bi-directionally coupled controller (or more precisely, each vehicle considers all others in its' control algorithm).
Similar to the speed injection attack, injecting false acceleration data causes attack impact dependent on the injected value; the attack impact is significant for positive acceleration values.
However, here the impact of false negative acceleration is also significant\,---\,however, at some point, the distance sensor notices that the distance is too large, and it resets to ACC behavior.

\section{Discussion}
\label{sec:discussion}
%

Our results show that most CACC controllers are unreliable when subject to effective denial of service attacks, and that the impact of injection is also significant for any data that is used in the controller directly.
This confirms earlier results on individual jamming and injection experiments, and shows that these results also hold when simulation parameters are changed.
Beyond this, our results show that the window of opportunity for the attacker to jam or transmit false information is broad, i.e., the attacker does not need to hit precisely the correct timing for the attack to work.
Our results can also be used as a guide for the attacker to further hone the effectiveness of an attack against more resilient control algorithms.

Overall, we can conclude that the Ploeg controller deals well with jamming even considering that this controller was only intended for intermittent, unreliable channels, and not for complete failure.
However, we still managed to cause some accidents even with this algorithm, suggesting that an eventual complete fallback to ACC may be advisable.
The consensus controller behaves differently, and only fails with a later starting point when jamming\,---\,future work could more thoroughly investigate the cause of this behavior.
Finally, we conclude that the constant spacing controller is completely unsuitable for unreliable communication; the window for causing a crash is at least 2 seconds.

For data injection attacks, the actual value transmitted by the attacker had an important influence over the attack impact, but the target speed of the leader did not matter much in comparison.
We expected our attacks to increase in impact as the overall platoon speed increases, because we expected the controllers to react more quickly in those scenarios.
Although this is the case for some position errors, the crash impact remains relatively consistent for the CACC controller.
For the Ploeg controller, the impact is dependent on the speed, but this dependence is mainly due to the speed-dependent gap that should occur between vehicles.
The one exception to all these observations is the consensus controller, which is surprisingly resilient against attacks, showing no impact from speed and acceleration falsification.
However, the impact of position shift on this controller is significant.
Nevertheless, this suggests that the consensus controller might be a good starting point for building a resilient controller, despite its' relative ineffectiveness, illustrated by a higher overall position error.

%
%

Our results are likely to extend to similar scenarios; the CACC and Ploeg controllers both rely only on the leader and the preceding vehicle, so any platoon long enough to contain the crashing vehicle is affected by potential crashes.
It is unclear how well the results transfer for the consensus controller, since its' intended behavior is quite different from the other two controllers, considering all the vehicles in the platoon for the control of each vehicle.
However, the overall methodology we applied here is not constrained to the scenario\,---\,the tools we provide can easily be used to further tune the parameters of the controllers for increased safety, or the attackers' parameters to improve attack effectiveness.

\section{Conclusion \& Future Work}
\label{sec:conclusion}
In this paper, we have discussed a more holistic attacker model, implemented several attacks and tested their impact on existing controllers using our proposed analysis methodology.
Overall, the results show not only that controllers are vulnerable to attack, but that there are significant differences between controllers in terms of how they are affected by attacks.
The research community should critically discuss the limitations of CACC under the influence of these attacks.
Our future work focuses on improving the attacks and studying their detection; however, detection alone is likely insufficient, because of the volume of crashes we encountered.
Work should therefore focus also on prevention, through the design of resilient controllers on the one hand, and effective detection and revocation mechanisms on the other.

Future work in false data injection could look at how other types of controllers behave, or study the effectiveness of simultaneously falsifying multiple fields in the beacon.
The latter is particularly relevant, because it circumvents many typical misbehavior detection mechanisms (which could detect most of the attacks we have described with reasonable accuracy, in part due to the values we chose).
An alternative route that has not yet been studied very well is the injection of false data into external sensors.
Some research has shown that camera and LIDAR sensors can be tricked easily in a lab environment~\cite{Petit2015}.
Once Plexe has been extended to include some sensor error models, experimenting with these attacks and combining them with false data injection will likely be an interesting path.
Even though attacks on sensors would rely on line-of-sight and additional equipment (i.e., it cannot be done by manipulating a vehicle with malware), we have also found that some injection attacks cause crashes between vehicles further behind the attacker.

For the development of future control algorithms, we suggest that authors consider a fallback mechanism similar to the Ploeg controller.
However, we suggest that this eventually downgrades from degraded CACC to pure ACC, because the error in the estimated acceleration will increase over time if a jamming attack is maintained long enough.
To improve reliability, using all available inputs may be a feasible approach to reduce the impact of attacks (e.g., as partially demonstrated by the consensus controller), especially because common controllers typically only use leader and preceding vehicle information.
This may not increase overall security, but will make it more difficult to successfully fine-tune attacks and achieve a desired impact.
Should the information show inconsistencies, a fallback mechanism could again be considered; here the challenge is ensuring that the entire platoon mutually decides on performing a fallback, to avoid a heterogeneous controller environment.

The attacks discussed in this work may be detected with relative ease using plausibility checks; however, the main aim of this work was to show how attack impact may be maximized.
Future work may look at the maximum tolerance of the attacks we designed; however, our currently proposed metrics are bound to the behavior of the leader.
The next step is then to decide how to use this information, apart from invalidating the vehicle in the SCMS~\cite{Whyte-SCMS}, preventing damage from ongoing attacks is also a goal.
If using a controller with graceful degradation (similar to Ploeg), the detection may trigger this functionality: when combined with effective revocation, this will ensure that temporarily degraded service is the best possible attack.
An interesting side aspect is that not all detection algorithms are compatible with platooning, because the corresponding behavior is different: it may thus be a feasible path to design specialized misbehavior detection mechanisms for the CACC scenario specifically.


\section*{Acknowledgment}
The authors would like to thank Johannes Diebold for the thesis upon which this work expands, and Henning Kopp for proof reading the paper.
Experiments for this work were performed on the computational resource bwUniCluster funded by the Ministry of Science, Research and the Arts Baden-W\"{u}rttemberg and the Universities of the State of Baden-W\"{u}rttemberg, Germany, within the framework program bwHPC.
This work was supported in part by the Baden-W\"{u}rttemberg Stiftung gGmbH Stuttgart as part of the project IKT-05 AutoDetect of its IT security research programme.

\balance



\bibliographystyle{IEEEtran}
\bibliography{IEEEabrv,references}

\begin{thebibliography}{10}
\providecommand{\url}[1]{#1}
\csname url@samestyle\endcsname
\providecommand{\newblock}{\relax}
\providecommand{\bibinfo}[2]{#2}
\providecommand{\BIBentrySTDinterwordspacing}{\spaceskip=0pt\relax}
\providecommand{\BIBentryALTinterwordstretchfactor}{4}
\providecommand{\BIBentryALTinterwordspacing}{\spaceskip=\fontdimen2\font plus
\BIBentryALTinterwordstretchfactor\fontdimen3\font minus
  \fontdimen4\font\relax}
\providecommand{\BIBforeignlanguage}[2]{{%
\expandafter\ifx\csname l@#1\endcsname\relax
\typeout{** WARNING: IEEEtran.bst: No hyphenation pattern has been}%
\typeout{** loaded for the language `#1'. Using the pattern for}%
\typeout{** the default language instead.}%
\else
\language=\csname l@#1\endcsname
\fi
#2}}
\providecommand{\BIBdecl}{\relax}
\BIBdecl

\bibitem{CSUR-Platooning}
D.~Jia, K.~Lu, J.~Wang, X.~Zhang, and X.~Shen, ``A survey on platoon-based
  vehicular cyber-physical systems,'' \emph{IEEE Communications Surveys
  Tutorials}, vol.~18, no.~1, pp. 263--284, Firstquarter 2016.

\bibitem{Rajamani-CC-book}
R.~Rajamani, \emph{Vehicle Dynamics and Control}.\hskip 1em plus 0.5em minus
  0.4em\relax Springer US, 2012.

\bibitem{RayaSecuritySurvey}
M.~Raya and J.-P. Hubaux, ``{Securing vehicular ad hoc networks},''
  \emph{Journal of Computer Security}, vol.~15, pp. 39--68, 2007.

\bibitem{Dadras-Adversarial}
S.~Dadras, R.~M. Gerdes, and R.~Sharma, ``Vehicular platooning in an
  adversarial environment,'' in \emph{Proceedings of the 10th ACM Symposium on
  Information, Computer and Communications Security}, ser. ASIA CCS '15.\hskip
  1em plus 0.5em minus 0.4em\relax New York, NY, USA: ACM, 2015, pp. 167--178.

\bibitem{DeBruhl-Misbehavior}
\BIBentryALTinterwordspacing
B.~DeBruhl, S.~Weerakkody, B.~Sinopoli, and P.~Tague, ``Is your commute driving
  you crazy?: A study of misbehavior in vehicular platoons,'' in
  \emph{Proceedings of the 8th ACM Conference on Security \& Privacy in
  Wireless and Mobile Networks}, ser. WiSec '15.\hskip 1em plus 0.5em minus
  0.4em\relax New York, NY, USA: ACM, 2015, pp. 22:1--22:11. [Online].
  Available: \url{http://doi.acm.org/10.1145/2766498.2766505}
\BIBentrySTDinterwordspacing

\bibitem{Amoozadeh-Commag}
M.~Amoozadeh, A.~Raghuramu, C.~n.~Chuah, D.~Ghosal, H.~M. Zhang, J.~Rowe, and
  K.~Levitt, ``Security vulnerabilities of connected vehicle streams and their
  impact on cooperative driving,'' \emph{IEEE Communications Magazine},
  vol.~53, no.~6, pp. 126--132, June 2015.

\bibitem{Alipour-Jamming}
A.~Alipour-Fanid, M.~Dabaghchian, H.~Zhang, and K.~Zeng, ``String stability
  analysis of cooperative adaptive cruise control under jamming attacks,'' in
  \emph{2017 IEEE 18th International Symposium on High Assurance Systems
  Engineering (HASE)}, Jan 2017, pp. 157--162.

\bibitem{Segata-Plexe}
M.~Segata, S.~Joerer, B.~Bloessl, C.~Sommer, F.~Dressler, and R.~Lo~Cigno,
  ``{PLEXE: A Platooning Extension for Veins},'' in \emph{6th IEEE Vehicular
  Networking Conference (VNC 2014)}.\hskip 1em plus 0.5em minus 0.4em\relax
  Paderborn, Germany: IEEE, December 2014, pp. 53--60.

\bibitem{Ploeg-Platooning}
J.~Ploeg, E.~Semsar-Kazerooni, G.~Lijster, N.~van~de Wouw, and H.~Nijmeijer,
  ``Graceful degradation of cacc performance subject to unreliable wireless
  communication,'' in \emph{Intl. IEEE Conference on Intelligent Transportation
  Systems (ITSC 2013)}, Oct 2013, pp. 1210--1216.

\bibitem{diBernardo-Consensus}
M.~di~Bernardo, A.~Salvi, and S.~Santini, ``Distributed consensus strategy for
  platooning of vehicles in the presence of time-varying heterogeneous
  communication delays,'' \emph{IEEE Transactions on Intelligent Transportation
  Systems}, vol.~16, no.~1, pp. 102--112, Feb 2015.

\bibitem{vanderHeijden2016}
R.~W. van~der Heijden, F.~Kargl, O.~M.~F. Abu-Sharkh, and A.~Al-Momani,
  ``Enhanced position verification for vanets using subjective logic,'' in
  \emph{Vehicular Technology Conference Fall}.\hskip 1em plus 0.5em minus
  0.4em\relax IEEE, 2016.

\bibitem{Gerdes-Efficiency}
R.~M. Gerdes, C.~Winstead, and K.~Heaslip, ``{CPS}: An efficiency-motivated
  attack against autonomous vehicular transportation,'' in \emph{Proc. Annual
  Computer Security Applications Conference}, ser. ACSAC '13.\hskip 1em plus
  0.5em minus 0.4em\relax New York, NY, USA: ACM, 2013, pp. 99--108.

\bibitem{Petit2015}
J.~Petit and S.~E. Shladover, ``Potential {Cyberattacks} on {Automated}
  {Vehicles},'' \emph{IEEE Transactions on Intelligent Transportation Systems},
  vol.~16, no.~2, pp. 546--556, Apr. 2015.

\bibitem{Feiri-CertOmission}
M.~Feiri, J.~Petit, and F.~Kargl, ``Congestion-based certificate omission in
  vanets,'' in \emph{Intl. Workshop on Vehicular Inter-networking, Systems, and
  Applications}, ser. VANET '12.\hskip 1em plus 0.5em minus 0.4em\relax NY,
  USA: ACM, 2012, pp. 135--138.

\bibitem{Douceur-Sybil}
J.~R. Douceur, ``The {Sybil} {Attack},'' in \emph{Peer-to-{Peer} {Systems}},
  ser. Lecture {Notes} in {Computer} {Science}, P.~Druschel, F.~Kaashoek, and
  A.~Rowstron, Eds., vol. 2429.\hskip 1em plus 0.5em minus 0.4em\relax Springer
  Berlin Heidelberg, Mar. 2002, pp. 251--260.

\bibitem{Joksch1993-Velocity}
H.~C. Joksch, ``Velocity change and fatality risk in a crash -- a rule of
  thumb,'' \emph{Accident Analysis \& Prevention}, vol.~25, no.~1, pp.
  103--104, 1993.

\bibitem{Larburu2010-Safe}
M.~Larburu, J.~Sanchez, and D.~J. Rodriguez, ``Safe road trains for
  environment: Human factors' aspects in dual mode transport systems,'' in
  \emph{ITS World Congress, Busan, Korea}, 2010.

\bibitem{Whyte-SCMS}
W.~Whyte, A.~Weimerskirch, V.~Kumar, and T.~Hehn, ``A security credential
  management system for v2v communications,'' in \emph{2013 IEEE Vehicular
  Networking Conference}, Dec 2013, pp. 1--8.

\end{thebibliography}
%
%
%

%

\end{document}